\documentclass[]{tPHM2e}
\usepackage{amsmath}
\usepackage{amssymb}
\usepackage{times}
\usepackage[varg]{txfonts}
\usepackage{color}

\doi{10.1080/14786435.20xx.xxxxxx}
\issn{1478-6443}
\issnp{1478-6435}
\jvol{00} \jnum{00} \jyear{2010} 
\date{Today}
\begin{document}
\title { 
On the Boson-Fermion resonant model on a lattice}
\author{{\bf R. Micnas} \\\vspace{6pt}
 Solid State Theory Division, Faculty of Physics,  A. Mickiewicz University, 
Umultowska 85, 61-614 Pozna\'{n}, Poland}
\maketitle
\begin{abstract}

We review briefly the  properties of  a mixture of mutually interacting bosons  
(bound  electron pairs) and itinerant fermions on a lattice (the boson-fermion
model).  The calculations of the superconducting phase transition temperature
($T_{c}$) and the phase diagram are the main concern. 
The self-consistent $T$-matrix method is applied to determine the
superconducting critical temperature from a pseudogap phase. 
The method takes into account the pairing
fluctuations 
effects. The $T$-matrix results for
$T_{c}$ are given for a 3D cubic lattice with tight-binding dispersion of
electrons and standard bosons, and they are also compared with those of the BCS-
mean-field approximation (MFA).  Our results describe the BCS-Bose-Einstein
condensation (BEC) crossover in the boson-fermion mixture with resonant
interaction. The energy scales involved in  the pseudogap formation are also
analysed. 

\end{abstract}
\date{\today}
\maketitle

\section{Introduction}
The scenario of coexisting local pairs (LPs) and itinerant electrons (a mixture
of charged $2e$ bosons and fermions), i.e. the boson-fermion model, for
non-conventional superconductors, chalcogenide glasses, and systems with
alternating valence,  was introduced about two decades ago
\cite{{beg},{rob87},{Micnas90}}. Because of the intersubsystem charge exchange
coupling,  an induced (resonant) pairing  mechanism can be active in this
model, which prompts  the superconducting state involving both 
boson and fermion
components. A related resonance-boson model of superconductivity was developed in
Refs.\cite{lee89}. A mixture of interacting  bosons and itinerant electrons  can 
display superconducting  characteristics which are intermediate between those
of the local pair superconductors and those of the homogeneous  BCS systems
\cite{{rob87},{Micnas90},{lee89},{jrmr96}}. 
The relevance of this two-component 
model for high temperature superconductors (HTS) and other short-coherence
length superconductors has been the concern of many authors
\cite{{rob87},Micnas90,lee89,jrmr96,NATO, RMunpub,
kostyrko,TDomanski,HCmodel,
larkin97,rome00,castro01,altman02,tsai06,prb21,rmetal03,
2rev,RM07,Mihlin09,Tadek,QMC,{levinR}}. 
It has also been applied as the two-channel model for description of
the BCS--BEC crossover from the atom Cooper pairs
to molecules in ultra-cold fermionic atomic gases with a Feshbach resonance
{\cite{{holland},{ohashi},{levinR}}.

Even though for HTS the 
boson-fermion (BF) model has been proposed
phenomenologically, it can also be obtained as an effective low-energy model
from microscopic formulation. 
For instance, within the polaron scenario, it has been derived from  the 
generalized periodic Anderson model
with on-site hybridization of wide- and narrow-band electrons, in which the
narrow-band electrons are locally strongly coupled with the lattice deformation \cite{rob87}.
In this context, LPs (bipolarons) are formed which are 
hard-core (charged $2e$) bosons made up of two tightly bound
fermions. 
Next, the plaquette BF model, an effective model for hole pairing in cuprates, has  been obtained
from the strongly correlated  Hubbard  model on the square lattice by the
contractor renormalization  method \cite{altman02}.  Several authors have
considered  the BF scenarios in the investigations 
of  superconductivity mechanism, exploring 
heterogeneity of the electronic structure of cuprate HTS, 
especially in the pseudogap phase,  
either in the momentum space (the Fermi arcs model)
\cite{larkin97,rome00,prb21,rmetal03,2rev,QMC}}  
or
in the real space  (charge and
spin inhomogeneities) \cite{castro01,tsai06}. 

For a review of two-component scenarios for non-conventional (exotic)
superconductors, see Ref. \cite{2rev}. Also, disorder and
inhomogeneity effects have been  recently  investigated in the (hard core)
boson-fermion model for non-conventional superconductors \cite{disorder1,
disorder2}. 

Thus, the boson-fermion model with resonant interaction is the basic model for 
superconductivity that has been adopted to explain high-Tc superconductivity 
and the BCS-BEC crossover in ultra-cold fermionic atomic gases
\cite{rob87,Micnas90,lee89,jrmr96,NATO,
RMunpub, kostyrko, TDomanski,HCmodel,
larkin97,rome00,castro01,altman02,tsai06,prb21,rmetal03,2rev,RM07,Mihlin09,
Tadek,QMC,levinR,holland,ohashi,
disorder1, disorder2, Coulomb,ELM,Yildirim,MItr,CrossoverSI,mapping}.

The purpose of this paper is analysis of the phase diagrams of the BF model on a lattice
and detail evaluation of the superconducting  transition temperature 
beyond BCS-MFA. 
We extend our previous study \cite{RM07}  based on a 
generalized $T$-matrix approach 
adapted to the BF model 
and provide further results.  
Our method explores the pairing
fluctuation theory of the BCS-BEC crossover for
single-channel fermion systems with attraction
\cite{zurich,levin,levinR, RM2001, ACRM14}. 
In Sec.2 we briefly outline the $T$-matrix formalism for the BF model 
and present derivation of equations which
determine $T_{c}$ from the pseudogap phase,
not described in Ref.\cite{RM07}. The numerical solutions to
these equations  for a simple cubic lattice, with the tight-binding dispersion for
fermions, are reviewed in Sec. 3.

\section{$T$-matrix formalism: Equations for 
 $T_{c}$ in the Boson--Fermion model}
We will consider the  boson-fermion model on a lattice 
described by the
following Hamiltonian \cite{RM07}:
\begin{eqnarray}\label{obmodel}
{\cal {H}} = \sum_{\mathbf{k}\sigma} (\varepsilon_{\mathbf{k}}-\mu) c^{\dagger}_{\mathbf {k}\sigma}
c_{\mathbf {k}\sigma} + \sum_{\mathbf {q}}(E^{0}_{\mathbf q}+2\Delta_{B}-2\mu)
b^{\dagger}_{\mathbf {q}}b_{\mathbf {q}} \nonumber \\
-\frac{U}{N}\sum_{\mathbf {k},\mathbf {k}',\mathbf {q}}
c^{\dagger}_{\mathbf {k}+\mathbf {q}/2,\uparrow} c^{\dagger}_{-\mathbf {k}+\mathbf {q}/2,\downarrow}
c_{-\mathbf {k}'+\mathbf {q}/2,\downarrow} c_{\mathbf{k}'+\mathbf {q}/2,\uparrow} \nonumber \\
+\frac{I}{\sqrt{N}}\sum_{\mathbf {q}}
\left(b^{\dagger}_{\mathbf {q}}B_{\mathbf {q}}+
B^{\dagger}_{\mathbf {q}}b_{\mathbf {q}}\right),
\end{eqnarray}
\begin{eqnarray} 
B^{\dagger}_{\mathbf q}=\sum_{\mathbf k}c^{\dagger}_{{\mathbf k}+{\mathbf
q}/2,\uparrow}c^{\dagger}_{-{\mathbf k} +{\mathbf q}/2,\downarrow}~,
\end{eqnarray}
where, $\sigma=\{\uparrow,\downarrow\}$,
$c^{\dagger}_{\mathbf {k}\sigma}$ and $c_{\mathbf {k}\sigma}$ 
are the fermion creation (annihilation) operators 
with momentum ${\mathbf k}$ and spin ${\sigma}$.
$b_{\mathbf q},b^{\dagger}_{\mathbf q}$ represent the 
boson operators satisfying the standard commutation relations:
$[b_{\mathbf q},b^{\dagger}_{\mathbf q'}]=\delta_{{\mathbf q}{\mathbf q'}},
[b_{\mathbf q},b_{\mathbf q'}]=0=[b^{\dagger}_{\mathbf q},b^{\dagger}_{\mathbf q'}]$.
$B^{\dagger}_{\mathbf q}$ stands for the singlet pair creation
operator of $c$-electrons.
$\varepsilon_{\mathbf k}$ is the electron band energy, 
$E^{0}_{\mathbf q}$ is the boson kinetic energy,  
and they are both
defined on the hypercubic lattice.  $E^{0}_{\mathbf 0}=0$.
$2\Delta_{B}$ is the bottom of the 
boson band and $\mu$ is the chemical potential. 
$I$ is the intersubsystem (resonant) coupling constant.
$U$ - the direct (non-resonant)
interaction between fermions. 
The total number of particles  per site is  $n=n_{F}+2n_{B}$,   where
$n_{F}=\frac{1}{N}\sum_{\mathbf {k}\sigma} \langle
c^{\dagger}_{\mathbf{k}\sigma}c_{\mathbf {k}\sigma}\rangle$ is the electron
concentration  and $n_{B}=\frac{1}{N}\sum_{\mathbf q}\langle
b^{\dagger}_{{\mathbf q}}b_{\mathbf q}\rangle$ - the boson concentration, $N$ -
the number of lattice sites.\\ The inherent property of the model is the
presence of pair exchange interaction ($I$) (or interconversion term), i.e. when
a boson is created ($b^{\dagger}_{{\mathbf q}}$)  simultaneously  a singlet-pair
of c-electrons is annihilated ($B_{\mathbf q}  $) and vice versa. If $I=0$, we
have two subsystems decoupled from each other  and they can undergo a
transition at $T_{BCS}$ (at weak $U$, for fermions) and $T_{BEC}$ (for
bosons). However, if the intersubsystem interaction $I\neq 0$,  
one common transition to the superfluid state will occur.\\  At first we consider the case of the
absence of the direct fermion interaction,  i.e. the case of $U=0$.

In the self-consistent $T$-matrix  approximation the fermionic ($G(k)$) and bosonic ($D(q)$) Green's
functions (GF) in the normal state satisfy the 
equations
\cite{jrmr96,RMunpub}:
\begin{eqnarray}\label{Dyson}
G(k) = \frac{1}{G_{0}^{-1}(k)-\Sigma_{F}(k)}~,\\\label{Fermionicself}
\Sigma_{F}(k) =\sum_{q}\Gamma(q)G(q-k)~,\\\label{Tmat}
\Gamma(q) =  I^2 D(q)~,\\\label{bosongf}
D(q) = \frac{1}{D^{-1}_{0}(q) - \Sigma_{B}(q)}~,\\\label{bosonself}
\Sigma_{B}(q) = -I^2 \Pi(q) ~,\\\label{psusceptibility}
\Pi(q) =\sum_{k}G(k)G(q-k)~,
\end{eqnarray}
where we used the four-vector notation: $k=({\mathbf k}, i\omega_{n}),~ q=({\mathbf q}, i\nu_{m})$, 
$ \sum_{k}= \frac{1}{\beta N}\sum_{{\mathbf k}}\sum_{\omega_{n}}$,
$ \sum_{q}= \frac{1}{\beta N}\sum_{{\mathbf q}}\sum_{\nu_{m}}$.  
$\omega_{n}=(2n+1)\pi/\beta$ and $\nu_{m}=2m\pi/\beta$ are the  
fermionic and bosonic Matsubara frequencies, respectively. $\beta=1/k_{B}T$.
The free fermionic GF is  
$G_{0}^{-1}(k)= i\omega_{n}-{\bar \varepsilon_{\mathbf k}}$,
while  the free bosonic GF:  $D^{-1}_{0}(q)= i\nu_{m} -2{\bar \Delta_{B}} -E^{0}_{\mathbf q}$.
${\bar \varepsilon}_{\mathbf k}=\varepsilon_{\mathbf k}-\mu$, 
$\bar{\Delta}_{B}=\Delta_{B}-\mu$. 
$\Sigma_{F}(k)$ and $\Sigma_{B}(q)$ are the fermion and boson self-energies,
respectively. $\Pi(q)$ is the pair susceptibility and $\Gamma(q)$ - the $T$-matrix.

The basic idea in the calculations of the critical temperature consists in the
following approximation for the fermionic self-energy $\Sigma_{F}(k)$.
We consider only  slow fluctuations of the pairing field and  
neglect temporal and spatial variations, i.e.  
the terms with $q\approx 0$ are the dominant ones in Eq.(\ref{Fermionicself})
\cite{{levin},{schmid},{tchernyshyov}}
\begin{eqnarray}
\Sigma_{F}(k) \approx G_{0}(-k)\sum_{q}\Gamma(q)
= \frac{\Delta_{pg}^2}{i\omega_{n}+{\bar \varepsilon_{-{\mathbf k}}}}\,.
\end{eqnarray}
\begin{eqnarray}\label{pg}
\Delta_{pg}^2=-\sum_{q}\Gamma(q)=-\frac{1}{\pi N}\sum_{\mathbf q}\int_{-\infty}^{\infty}
Im\Gamma({\mathbf q},\Omega)b(\Omega)d\Omega \, ,
\end{eqnarray}
$b(\Omega)=1/\left[\exp{(\beta\Omega)}-1\right]$ is the Bose function.
Eq.(\ref{pg}) determines  a
pseudogap parameter $\Delta_{pg}$.
The fermionic GF (Eq.(\ref{Dyson})) takes the following form:
\begin{eqnarray}\label{BCSGF}
G(k)=\frac{i\omega_{n}+{\bar \varepsilon_{\mathbf k}}}
{(i\omega_{n})^2-E_{\mathbf k}^2}\;,\; E_{\mathbf k}=\
\sqrt{{\bar \varepsilon_{\mathbf k}}^2+\Delta_{pg}^2}\,,   
\end{eqnarray}
which is reminiscent of the standard BCS expression with the quasiparticle energy 
$E_{\mathbf {k}}$.
Using this GF, the pair susceptibility $\Pi(q)$ is calculated. In the following
instead of the full calculation of
$\Pi(q)$ \cite{{schmid},{tchernyshyov},{ACRM14}}  the partial dressing is adopted
which was considered in detail by Kadanoff and Martin \cite{kadanoff-martin} and
Levin et al. \cite{levin}
 \begin{eqnarray}
\Pi(q)\approx \sum_{k}G(k)G_{0}(q-k)~,
\end{eqnarray}
which  with the use of Eq.(\ref {BCSGF}) and after performing 
the summation over the Matsubara frequencies takes the form:

\begin{equation}\label{gg0}
\Pi(q)=-\frac{1}{N}\sum_{\mathbf k}\left[
\frac{f(E_{\mathbf k}) +f({\bar \varepsilon_{{\mathbf q}-{\mathbf k}}})-1}
{{\bar \varepsilon_{{\mathbf q}-{\mathbf k}}}+E_{\mathbf k}-i\nu_{m}}u_{\mathbf k}^2
+\frac{f({\bar \varepsilon_{{\mathbf q}-{\mathbf k}}})-f(E_{\mathbf k})}
{{\bar \varepsilon_{{\mathbf q}-{\mathbf k}}}-E_{\mathbf k}-i\nu_{m}}v_{\mathbf
k}^2 \right] \;,
\end{equation}

$u_{\mathbf k}^2+v_{\mathbf k}^2=1$,~~ $u_{\mathbf k}^2=\frac{1}{2}(1+
\frac{{\bar \varepsilon_{\mathbf k}}}{E_{\mathbf k}})$, 
$f(\omega)=1/\left[\exp{(\beta\omega)}+1\right]$ is the Fermi function.

In the single channel 
fermion system the $T$-matrix is given by 
$\Gamma^{-1}(q)=g^{-1}+\Pi(q)$, where $g$ is the coupling constant. Thus, the
condition $\Gamma^{-1}(0)=0$ determines $T_{c}$ and we obtain  the system of
three coupled equations for $\Delta_{pg}$, $T_{c}$ and $\mu$, which were 
studied in detail for the continuum and lattice fermions with attractive 
interaction in Refs.\cite{levinR,levin,ACRM14}. We call this $T$-matrix approach as
the $(GG_{0})G_{0}$ scheme, which indicates the way the Greens functions enter
the pair susceptibility and the  fermion self-energy.
 
In our boson-fermion model we have to consider 
the  bosonic GF Eq.(\ref{bosongf}), which is given by
\begin{eqnarray}\label{Tmatrix}
\Gamma(q) = I^2 D(q)=
\frac{I^2}
{i\nu_{m} -2{\bar \Delta_{B}} -E^{0}_{\mathbf q}+I^2 \Pi(q)}\,.
\end{eqnarray}
The divergence of the generalized $T$-matrix $\Gamma(q)$ at $q=0$, i.e. $({\mathbf q}=0, \Omega=0)$ is the same as for the
bosonic  GF $D(q)$ and yields the equation for $T_{c}$:
\begin{eqnarray}\label{crit}
2{\bar \Delta_{B}} +E^{0}_{\mathbf 0}-I^2 \Pi(0)=0\;,
\end{eqnarray}
\begin{eqnarray}\label{crittemp}
2{\bar \Delta_{B}}+E^{0}_{\mathbf 0}=
\frac{I^2}{N}\sum_{\mathbf k}\frac{\tanh(\beta_{c}E_{\mathbf k}/2)}{2E_{\mathbf k}}~,
\end{eqnarray}
where we used Eq.(\ref{gg0}): 
$\Pi(0)=\Pi({\mathbf {0}},0;T_{c})=\frac{1}{N}\sum_{\mathbf k}\frac{\tanh(\beta_{c}E_{\mathbf k}/2)}{2E_{\mathbf k}}~.$
The number of fermions is given by:
\begin{equation}\label{fermionnumber}
n_{F}=\frac{2}{\beta N}\sum_{{\mathbf k},\, \omega_{n}}e^{i\omega_{n}\eta}G(k)=
\frac{1}{N}\sum_{\mathbf k}\left[1-\frac{{\bar \varepsilon_{\mathbf k}}}{E_{\mathbf
k}}\tanh{(\beta_{c}E_{\mathbf k}/2)}\right] \;.
\end{equation}
The number of bosons is:
\begin{equation}\label{bosonnumber}
n_{B}=-\frac{1}{\beta N}\sum_{{\mathbf q},\, \nu_{m}}e^{i\nu_{m}\eta}D(q)=
-{\frac{1}{\pi N}\sum_{\mathbf q}}\int_{-\infty}^{\infty}
ImD({\mathbf q},\Omega)b(\Omega)d\Omega \;.
\end{equation}
($\eta=0^{+}$).
The total number of particles in the system $n$ is conserved 
\begin{eqnarray}\label{numberconserv}
n=n_{F}+2n_{B}\;.
\end{eqnarray}
Thus, by comparing Eq.(\ref{pg}) and Eq.(\ref{bosonnumber}) 
one gets that at $T_{c}$:
\begin{eqnarray}\label{relat}
\Delta_{pg}^2=I^2n_{B}\;.
\end{eqnarray}
The above system of self-consistent equations
(\ref{pg},\ref{crittemp},\ref{numberconserv},\ref{relat})
determines $\Delta_{pg}$, $T_{c}$ and the chemical potential $\mu$.
In comparison with  BCS-MFA, 
we have taken  into account the boson self-energy effect and included pairing
fluctuations.

Next, we examine the  direct interaction between fermions. The effect of $U$
will be included in the Random Phase Approximation - like method, just
treating it in the ladder approximation \cite{kostyrko,ohashi}. This amounts to
generalization of the $T$-matrix as  follows:
\begin{eqnarray}
\Gamma(q)=\frac{-U_{eff}(q)}{1-U_{eff}(q)\Pi(q)}~,
\end{eqnarray}
where $U_{eff}(q)=U-I^2D_{0}(q)$ is the effective pairing interaction
and the bosonic self-energy is given by:
\begin{eqnarray}
\Sigma_{B}(q)=-I^2{\tilde \Pi}(q)~, ~
{\tilde \Pi}(q)=\frac{\Pi(q)}{1-U\Pi(q)}~.
\end{eqnarray}
The transition temperature is given by the Thouless criterion of the divergent
$T$-matrix, i.e. $1-U_{eff}(0)\Pi(0)=0$. Simultaneously one checks that 
the bosonic GF is divergent indicating a common transition in the
system. The pseudogap parameter and the number of bosons still satisfy the same
general equations (\ref{pg}) and (\ref{bosonnumber}).

Furthermore, in the numerical calculations of $T_{c}$, we use the following procedure to
determine the pseudogap parameter $\Delta_{pg}$. Such a procedure
should be reasonable for moderate to strong intersubsystem coupling. 
The analytically continued pair susceptibility 
(Eq.(\ref{gg0}))  has the expansion:
\begin{eqnarray}\label{expansion}
\Pi({\mathbf q},\Omega)-\Pi(0,0)\approx A_{0}'
(\Omega-\Omega_{{\mathbf q}})+iA_{0}''\Omega \;.
\end{eqnarray}
We make an assumption that:
$\epsilon \equiv A_{0}''/A_{0}'\ll1$~,
 and use te relation:
 $\lim_{\epsilon\rightarrow0}\frac{\epsilon}{x^2+\epsilon^2}=\pi\delta(x)$.
Using Eq.(\ref{pg}) and Eq.(\ref{crit}), the equation for  
pseudogap parameter takes the following  form at $T_{c}$ for $U=0$:

\begin{eqnarray}\label{PGparameter}
\Delta_{pg}^2=\frac{I^2}{1+I^2A_{0}'}{\frac{1}{N}}
\sum_{\mathbf q\neq 0}b\left(
\frac{E^{0}_{\mathbf q}+I^2A_{0}'\Omega_{\mathbf q}}
 {1+I^2A_{0}'}\right)~,
\end{eqnarray}
\begin{eqnarray}\label{pairdisp}
\Omega_{\mathbf q}=\frac{1}{A_{0}'}
\frac{1}{N}\sum_{\mathbf k}\left[ 
\frac{f(E_{\mathbf k})+f({\bar \varepsilon}_{{\mathbf q}-{\mathbf k}})-1}
{{\bar \varepsilon}_{{\mathbf q}-{\mathbf k}} +E_{\mathbf k}}u_{\mathbf k}^2 
+\frac{f({\bar \varepsilon}_{{\mathbf q}-{\mathbf k}})-f(E_{\mathbf k})}
{{\bar \varepsilon}_{{\mathbf q}-{\mathbf k}} -E_{\mathbf k}}v_{\mathbf k}^2 \right. \nonumber \\
+\left.\frac{1-2f(E_{\mathbf k})}{2E_{\mathbf k}}\right]\;,\;\;
\end{eqnarray}

\begin{eqnarray}\label{coeff}
A_{0}'=\frac{1}{2\Delta_{pg}^2}\left[n_{F}-
\frac{1}{N}\sum_{\mathbf k}2f(\bar \varepsilon_{\mathbf k})\right]~.
\end{eqnarray}
We point out that two  
kinds of (coupled) bosonic contributions are present
in $\Delta_{pg}$ (Eq.(\ref{PGparameter})). The one ($ \Omega_{\mathbf q} $) is
from long-lived pairs of $c$-electrons with finite ${\mathbf q}$ 
and the second  is
from the direct hopping of bosons $E_{\mathbf q}^{0}$. The pair dispersion
$\Omega_{{\mathbf q}}$ in the small ${\mathbf q}$
expansion is of the  form:
\begin{equation}
\Omega_{{\mathbf q}}=C{\mathbf q}^2=\frac{{\mathbf q}^2}{2M^{\star}}~,
\end{equation}
 $M^{\star}$ is the effective mass. 

For the case $I\neq 0$, $U\neq 0$ we obtain the following equation for $T_{c}$:
\begin{equation}\label{Tc(U)}
1=\left(U+\frac{I^2}{2(\Delta_{B}-\mu)}\right)\frac{1}{N}\sum_{\mathbf k}\frac{\tanh(\beta_{c}E_{\mathbf k}/2)}
{2E_{\mathbf k}}~.
\end{equation}
 With the expansion Eq.(\ref{expansion})  the pseudogap parameter and the number
 of bosons satisfy the following  equations:
\begin{eqnarray}
\Delta_{pg}^2=\frac{I_{eff}^2}{1+I_{eff}^2A_{0}'}{\frac{1}{N}}
\sum_{\mathbf q\neq 0}b\left(\frac{E^{0}_{\mathbf q}+I_{eff}^2A_{0}'\Omega_{\mathbf q}}
 {1+I_{eff}^2A_{0}'}\right)~~,\\ \nonumber
I_{eff}^2= (2{\bar \Delta_{B}}U+I^2)^2/I^2~~,
\end{eqnarray}
\begin{eqnarray}
n_{B}=\frac{1}{1+K_{eff}^2A_{0}'}{\frac{1}{N}}
\sum_{\mathbf q\neq 0}b\left(\frac{E^{0}_{\mathbf q}+K_{eff}^2A_{0}'\Omega_{\mathbf q}}
{1+K_{eff}^2A_{0}'}\right)~~,\\ \nonumber
K_{eff}^2=I^2/(1-U\Pi(0))^2~,
\end{eqnarray}
where $ \Omega_{\mathbf q}$ and  $A_{0}'$ are given by Eqs.(\ref{pairdisp}-\ref{coeff}) and
$\Pi(0)$ by 
\begin{eqnarray}\label{pi0}
\Pi(0)= \frac{1}{N}\sum_{\mathbf k}\frac{\tanh(\beta_{c}E_{\mathbf k}/2)}
{2E_{\mathbf k}}.
\end{eqnarray}
Using  Eq.(\ref{Tc(U)}) and expression for  
$I_{eff}^2$ one gets $K_{eff}^2=I_{eff}^2$, thus 
a similar relationship 
to that of 
(\ref{relat}) holds at $T_{c}$:
\begin{equation}
\Delta_{pg}^2=I_{eff}^2 n_{B}~.
\end{equation}
The above equations for $T_{c}$ and $\Delta_{pg}$ have to be solved together with the condition for
conservation of the total number of particles (\ref{numberconserv}), where $n_{F}$ is given by
Eq.(\ref{fermionnumber}). \cite{RM07}.

\section{Numerical results}
The numerical solutions to the equations for $T_{c}$  are  presented below  
for a simple cubic (sc)  lattice.
The electron band energy is given by:
$\varepsilon_{\mathbf k} = D (1-\gamma_{\mathbf k})$; $D=zt$, 
$\gamma_{\mathbf k}=\left[\cos(k_{x})+\cos(k_{y})+\cos(k_{z})\right]/3$,
where $t$ - nearest neighbour hopping parameter of c-electrons  and  $z$ -the
coordination number. For the kinetic energy of free bosons we take: 
$E^{0}_{\mathbf q}=J_{0}-J_{{\mathbf q}}$ 
$J_{\mathbf q}=J_{0}  \gamma_{\mathbf q},  ~J_{0}={\it z}J$,
$J$ -the direct boson hopping amplitude.
The momentum summations are over the first Brillouin zone. 
Furthermore, 
in the plots we will use the half of the electron bandwidth ($ D $) 
as an energy unit.

Figs.1-3 show the plots of $T_{c}$, $\Delta_{pg}(T_{c})$, 
and  $\mu(T_{c})$,
together with fractions of $n_{F}$ and $n_{B}$ 
versus the bosonic level position $\Delta_{B}$, across the BCS-BEC crossover. 
(See also Fig.1 in Ref. \cite{RM07}, for different $n$ and interaction
parameters).
%
The direct boson hopping (Figs.1,3) and
interaction between fermions are taken into regard (Figs.1-3). 
In Figs. (1,3) we set  $E^{0}_{\mathbf q}\simeq J {\mathbf q}^2$ and $J/t=1/2$ , this
corresponds to $ m_{B}=2m_{F}$, where $m_{B}=1/(2J)$, $m_{F}= 1/(2t)$ are (bare)
masses of bosons and fermions on the lattice, respectively,
($\hbar=a=1$, $a$ - the lattice spacing). 

As it is clearly seen from Fig.1 and 2 the superfluid phase transition changes 
in a smooth way from BCS-like to BEC-like
when the pairing correlations
are incorporated.
\begin{figure}
\begin{center}
\resizebox*{9.8cm}{!}
{\includegraphics{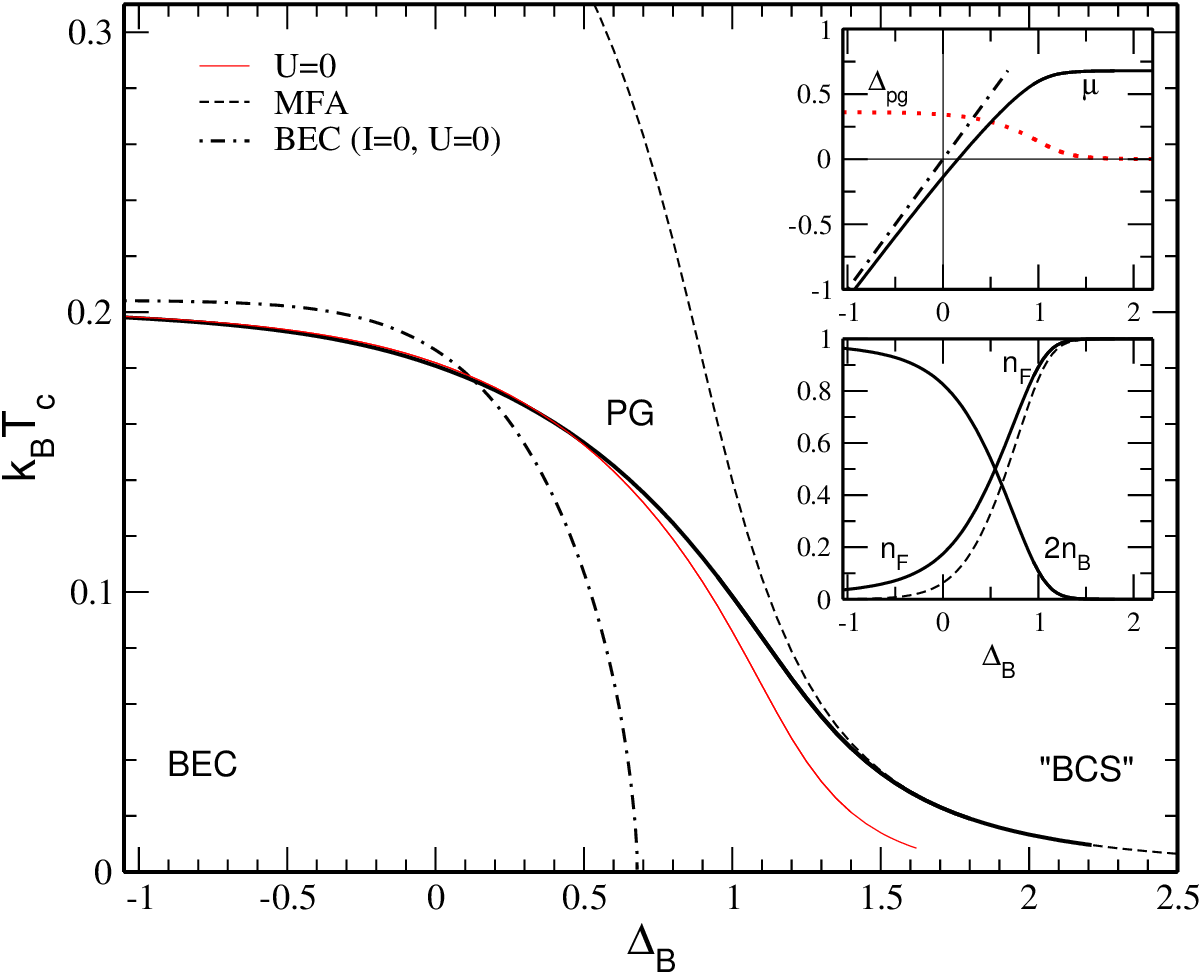}}
\caption{Computed self-consistent $T$-matrix results for the boson-fermion
model with resonant interaction (units $D=1$). $n=0.45$, $|I_{0}|/D=0.75$,
($I=-|I_{0}|$), $U/D=0.125$, $D=6t$. $J/t=0.5$. The superfluid
transition temperature obtained in $T$-matrix:  $T_{c}$ vs $\Delta_{B}$ is shown
by the solid curve; the thin red solid curve is for $U=0$.  The dashed curve --
the BCS-MFA transition temperature.  The dash-dotted curve --
the  $T_{BEC}$  for free boson-fermion mixture  
but with the constraint $n=n_{F}+2n_{B}$.  
Upper inset shows the variation of chemical potential $\mu(T_{c})$ (solid curve) 
and pseudogap parameter $\Delta_{pg}(T_{c})$ (red dotted curve) vs $\Delta_{B}$ ;  
the dash-dotted line -- the chemical potential for BEC
transition without interactions ($\mu=\Delta_{B}$). Lower inset:
fractions of $n_{F}$, $n_{B}$  and $n^{0}_{F}$ (dashed) vs $\Delta_{B}$ at $T_{c}$, normalized to the
total n, across the BCS-BEC crossover; 
$n^{0}_{F}$-fraction of unpaired fermions at $T_{c}$. See also Fig.1 in Ref.\cite{RM07}. }
\end{center}
\label{Fig1}
\end{figure}

Let us summarize the main features of three regimes of this evolution.\\
(i) The renormalized level  energy   (by the
boson self-energy) $\Delta^{*}_{B}$ is negative, i.e.
$2\Delta^{*}_{B}=2\Delta_{B}+\Sigma_{B}(q)<0$.  
In such boson predominant region, 
the bosons are essentially undamped (do not decay), 
the $\mu$ is negative, and for large negative $\Delta_{B}$, 
the $T_{c}$ approaches the $T_{BEC}$  for free bosons from below. 
What is more, the strong effective attractive interaction mediated by the
bosons leads to formation of preformed fermion pairs on the 
bosonic side of the crossover (compare $n_{F}$ and $n^{0}_{F}$ in lower inset 
in Fig.1).  \\
(ii) If $\Delta^{*}_{B}>0$, 
the interconversion  boson-pair of fermions (c-electrons) process
gives rise to the resonance superfluidity and to the elevation  
of $T_{c}$. 
In addition, the range
of resonant (or mixed) superfluidity is  associated with a pseudogap (PG).\\ 
(iii) At last, in the  BCS-like regime, predominant by fermions, the
boson fraction $n_{B}/n$ is small,  and $T_{c}$ approaches the BCS-MFA result.
In this case  
the chemical potential is very close to the Fermi energy and the pseudogap
becomes quite small (upper and lower inset in Fig.1).  Even a weak direct
attraction $U$ enlarges the BCS-like regime  (See Fig.1 for
$U/D=0.125$ and $U=0$, respectively),  but the repulsive $U$ reduces
it \cite{RM07}.

It is clear that with decreasing $J$, the BEC asymptote  to $T_{c}$ will be lower
because of larger $m_{B}$, however the BCS-like regime will be only little
affected because of small $n_B$. In consequence,  the smooth crossover plot of
$T_{c}$ will exhibit a round maximum inside the resonance regime for a definite
$J/t<0.5$.

\begin{figure}
\begin{center}
\resizebox{8cm}{!}
{\includegraphics{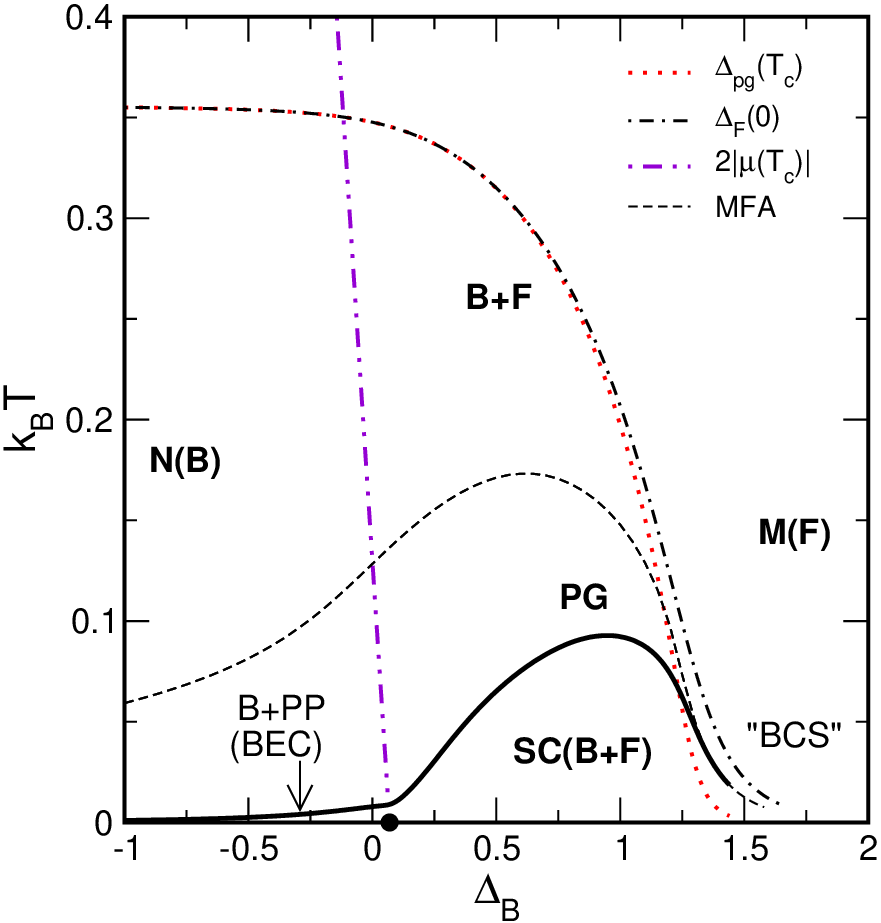}}
\caption{
Phase diagram of the boson-fermion model on a sc lattice with the direct boson
hopping suppressed (units $D=1$). $n=1$, $|I_{0}|/D=0.5$, $U/D=0.05$, $J=0$. The
solid line shows $T_{c}$ vs. $\Delta_{B}$  calculated from the self-consistent
$T$-matrix approach. The dash-double-dotted line is for the chemical potential
(from $T$-matrix) when $\mu$ is negative. B+PP - region of hybridized bosons and
preformed pairs of c-electrons.  SC(B+F) - resonance superconductivity phase.
M(F) - metallic phase dominated by fermions. PG - pseudogap region. N(B) -
normal phase dominated by bosons. Dot marks the crossover point in the
superconducting ground state. See text.}
\label{Fig2}
\end{center}
\end{figure}

Fig.2 shows the phase diagram of the BF model in the case when the direct boson
hopping is suppressed ($J=0$), i.e. the bosons are initially immobile, and 
$n=n_{F}+2 n_{B}=1$.  It has been obtained from $T$-matrix calculations 
supplemented by the analysis of superconducting ground state in BCS-MFA
\cite{RMunpub}. Let us stress that the presence of the resonant coupling ($I$)
alone is  sufficient to establish superfluidity in the BF mixture, with the
region of resonance (or mixed) superfluid. 
 In addition (Fig.2), the energy scale for the pseudogap obtained in the
$T$-matrix ($\Delta_{pg}(T_{c})$) is compared with the superconducting gap
parameter at $T=0K$ ($\Delta_{F}(0)$), determined in the BCS-MFA. We note that
these characteristic parameters are close to each other in the regime dominated
by bosons and differ in the BCS-like regime. The MFA $T_{c}$ (thin dashed line
in Fig.2), beyond the BCS-region, provides only the pairing scale for the
formation of incoherent   Cooper (fermion) pairs, and bounds a PG region.  The
dash-double-dotted line for $2\mu$ negative (being close to the molecule binding
energy) separates the bosonic regime. 
In the bosonic regime (B+PP) we have 
hybridized bosons and preformed fermion pairs, with the unique branch of
excitations given by the pole of bosonic GF $D(\mathbf q, \Omega)$.

In our numerical computations we assumed the parabolic spectrum for the bosons,
but keep the full tight-binding dispersion for the fermions. It was mainly
dictated by the simplifications due to 
the long-wave expansion of the pair susceptibility. 
For the parabolic boson dispersion we have a
simple result for BEC transition of free bosons 
$k_{B}T_{c}/2J=\frac{2\pi}{\zeta(3/2)^{2/3}} n_{B}^{2/3}=3.3125 n_{B}^{2/3}$.
Inclusion of the lattice dispersion for bosons $E^{0}_{\mathbf q}$,
is numerically possible, and  it results in  quantitative
improvement of $T_{c}$, in the BEC regime \cite{RMunpub}.

\subsection{Low-density and broad resonance case}
In this subsection we discuss the low density case. For moderate resonant coupling $I$ the
evolution of $T_{c}$ vs $\Delta_{B}$ is similar to that presented above. 
Particularly interesting case is  the limit of large $I$, large $\Delta_{B}$, 
for low density (in cold gases referred to as a broad resonance).  We observe
that in such a case ($I\rightarrow \infty, \Delta_{B} \rightarrow \infty$,  but
$ -\frac{I^2}{2\Delta_{B}}$ finite) the model effectively reduces to the single band
(or one-channel) fermion model with an effective attraction.  Fig.3 shows the
numerical results  for (relatively) large coupling $I$. In contrast to the
continuum model, with decreasing
$\Delta_{B}$ the critical temperature sharply decreases 
away from the unitarity (before reaching $T_{BEC}$), which is specific to the lattice model.
In Fig.3, in the fermion dominated regime, we find that  $T_{c}$ in the BF model
determined in the $T$-matrix approach, practically follows the behavior of  $T_{c}$ in the
attractive Hubbard model with an effective attraction  ${\bar
U}_{eff}=U+\frac{I^2}{2\Delta_{B}}$.

\begin{figure}
\begin{center}
\resizebox*{9cm}{!}{\includegraphics{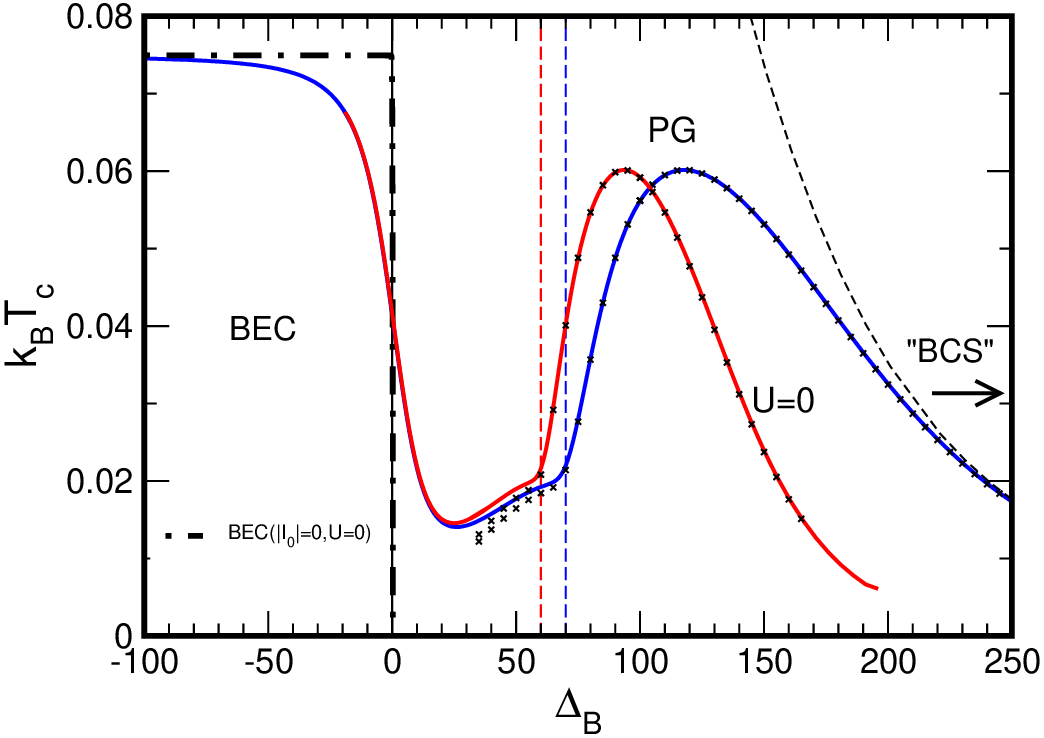}}
\caption{Evolution of $T_{c}$ from  BEC to BCS like limit calculated in the
$T$-matrix approach for large intersubsystem coupling and low density (units $D=1$). $n=0.1$,
$|I_{0}|/6t=15$, $U/6t=0.25$, $J/t=0.5$ $~(m_{B}=2m_{F})$. The denotations as in Fig.1.
 Crosses mark $T_{c}$ obtained in
$T$-matrix for the Hubbard model with an effective attraction ${\bar U}_{eff}$ and filling $n_{F}=0.1$. 
The dashed curve is for BCS-MFA.
Thin dashed vertical lines denote the
points for which the chemical potential passes zero, 
thus marking the beginning of a Bose regime.}
\end{center}
\label{Figlownb}
\end{figure}

\section{Outlook}
In summary,  we  have further investigated  the superfluid phase transition
temperature and the phase diagram of the boson-fermion model with resonant
interaction on the lattice.  
We also discussed, in terms of $T$-matrix
many-body formalism, the way the pseudogap physics
can be incorporated in  the description of boson-fermion system with resonant
interaction. The results obtained, describe mostly
the BCS-BEC crossover for $T_{c}$, with variable bosonic (LP) level position, 
and complement those of Ref.\cite{RM07}. The interesting region of the resonant
(or mixed) superconductivity is preceded by the pseudogap appearing because of 
pairing fluctuations. 

We considered  the BF system with standard lattice bosons, however the presented
approach can be extended to the important case of  hard-core bosons \cite{2rev,
RMunpub, RM07}. In addition, our study can be applied to: (i) the BF model with
resonant d-wave pairing on quasi 2D lattice \cite{RMunpub} (ii) description of
the BCS-BEC crossover in a neighbourhood of the quantum superfluid-band
insulator transition, which occurs in the model (\ref{obmodel}) for n=2
\cite{rob87, RMunpub},{\cite{CrossoverSI}}. Another extension concerns the application of the
$(GG)G_{0}$ $T$-matrix scheme to the BF model \cite{{RMunpub},ACRM14}.

The boson-fermion model we  studied 
can also be treated as
special
case of a 
general coupled boson-fermion-Hubbard model 
written in the Wannier basis:
\begin{eqnarray}
{\cal H} =   {\cal H}_B +{\cal H}_F+ {\cal H}_{1} ,
\end{eqnarray}
\begin{eqnarray}\label{Bosonpart}
{\cal H}_B = \sum_{i}(2\Delta_{B} - 2\mu)b^{\dagger}_{i} b_{i} - \sum_{i,j}J_{ij}
b^{\dagger}_{i} b_{j} + \frac{1}{2}U_B \sum_{i}n^{B}_{i}(n^{B}_{i}-1), \
\end{eqnarray}
\begin{eqnarray}\label{Fermionpart}
{\cal H}_F = \sum_{i,j,\sigma}t_{ij} c^{\dagger}_{i\sigma} c_{j\sigma} + 
\sum_{i,\sigma}(D - \mu)c^{\dagger}_{i\sigma}c_{i\sigma}+ 
\frac{1}{2}U_F \sum_{i}n^{F}_{i}(n^{F}_{i}-1),
\end{eqnarray}
\begin{eqnarray}
 {\cal H}_1= I\sum_{i} (b^{\dagger}_{i} c_{i\downarrow}c_{i\uparrow}+
c^{\dagger}_{i\uparrow}c^{\dagger}_{i\downarrow} b_{i}) + U_{BF}\sum_{i}n^{F}_{i}n^{B}_{i},
\end{eqnarray}

and $n = n_F + 2n_B.$\\
$n^{F}_{i}=n_{i\uparrow}+n_{i\downarrow}$, $n_{i\sigma}=c^{\dagger}_{i\sigma}c_{i\sigma}$,
$n^{B}_{i}=b^{\dagger}_{i} b_{i}$. $[b_{i},b^{\dagger}_{j}]=\delta_{ij}$. 
$\{c_{i\sigma},c^{\dagger}_{j\sigma'}\}=\delta_{ij}\delta_{\sigma\sigma'}$
$n_F=\frac{1}{N}\sum_{i\sigma}\langle n_{i\sigma}\rangle$, $n_B=\frac{1}{N}\sum_{i}\langle n^{B}_{i}\rangle$. 
$D=zt$.
The bosonic part ${\cal {H}}_{B}$ (\ref{Bosonpart}) is described by the  boson
Hubbard model  with the on-site repulsion $U_{B}$ and fermionic part ${\cal
{H}}_{F}$ (\ref{Fermionpart}) by the Hubbard model with the on-site interaction
$U_{F}$. The intersubsystem interactions are specified by the interconversion
term ($I$) and the boson-fermion repulsion ($U_{BF}$). \\
If $U_{B}=U_{BF}=0$, after transforming to ${\bf k}$ - space:
$c_{j\sigma}=\frac{1}{\sqrt N}\sum_{\mathbf k} 
e^{i {\mathbf k}\cdot{\mathbf R}_{j}}c_{\mathbf {k}\sigma},~
b_{j}=\frac{1}{\sqrt N}\sum_{\mathbf q} 
e^{i {\mathbf q}\cdot{\mathbf R}_{j}}b_{\mathbf q}$, 
we obtain the boson-fermion model (\ref{obmodel}), where $\varepsilon_{\mathbf{k}}$ 
($J_{\mathbf q})$ is
related to Fourier transform of $t_{ij}$ ($J_{ij}$), $\Delta_{B}\rightarrow
\Delta_{B}-J_{0}/2$ and $U_{F}=-U$. \\If $U_{B}\rightarrow \infty$, 
and keeping only the two lowest boson states, one gets the case of hard-core bosons (or pseudospins),
which satisfy the Pauli spin $1/2$ commutation relations (see Refs.\cite{rob87,RM07}). \\The above
general model is of interest for non-conventional superconductors as well as for boson-fermion
mixtures loaded in optical lattices.

\section{Acknowledgements}
I would like to thank  S. Robaszkiewicz for helpful discussions.

\end{document}